\documentclass[aps,pra,twocolumn,superscriptaddress,longbibliography]{revtex4-2}
\usepackage{amsfonts,amsmath,amssymb,epsfig}
\usepackage{mathtools}
\usepackage{epsfig}
\usepackage{graphicx}
\usepackage[dvipsnames]{xcolor}
\usepackage{color}
\usepackage{soul}
\usepackage{float}
\usepackage{soul}
\usepackage{epstopdf}
\usepackage{array}
\usepackage{tabularx}
\usepackage[utf8]{inputenc}
\usepackage[T1]{fontenc}
\usepackage[colorlinks,linkcolor=blue, citecolor=blue,
urlcolor=blue]{hyperref}
\usepackage{orcidlink}

\newcommand\eref[1]{(\ref{#1})}

\newcommand\fref[1]{Fig.~\ref{#1}}
\newcommand\sref[1]{Sec.~\ref{#1}}
\graphicspath{{./images/}}
\newcommand\revise[1]{#1}

\begin{document}
\title{Regimes of Steady-State Turbulence in a Quantum Fluid}
\author{Tommy Z. Fischer}
\affiliation{Department of Physics, University of Otago, Dunedin, New Zealand}
\affiliation{Dodd-Walls Centre for Photonic and Quantum Technologies}
\author{Ashton S. Bradley~\orcidlink{0000-0002-3027-3195}}
\affiliation{Department of Physics, University of Otago, Dunedin, New Zealand} 
\affiliation{Dodd-Walls Centre for Photonic and Quantum Technologies}

\date{\today}
\begin{abstract}
We simulate the Gross-Pitaevskii equation to model the development of turbulence in a quantum fluid confined by a cuboid box potential, and forced by shaking along one axis. We observe the development of isotropic turbulence from anisotropic forcing for a broad range of forcing amplitudes, and characterise the states through their Fourier spectra, vortex distributions, and spatial correlations. For weak forcing the steady-state wave-action spectrum exhibits a $k^{-3.5}$ scaling over wavenumber $k$; further decomposition uncovers the same power law in both compressible kinetic energy and quantum pressure, while the bulk superfluid remains phase coherent and free from extended vortices. As the forcing energy exceeds the chemical potential, extended vortices develop in the bulk, disrupting the $k^{-3.5}$ scaling. The spectrum then transitions to a $k^{-7/3}$ regime for compressible kinetic energy only, associated with dense vortex turbulence, and phase coherence limited to the healing length. The strong forcing regime is consistent with an inverse cascade of compressible energy driven by small-scale vortex annihilation. 
\end{abstract}
\flushbottom
\maketitle

\section{Introduction}
Turbulence is a ubiquitous phenomenon in nature, observed in a wide range of systems from the atmosphere to the oceans~\cite{grant_turbulence_1962}, and in the dynamics of quantum fluids~\cite{henn_emergence_2009,neely_characteristics_2013}.
Yet the turbulent flows understood in classical fluids~\cite{kolmogorov_local_1941} assume a different character in the quantum realm. The pursuit of understanding quantum turbulence in highly controllable Bose-Einstein condensates (BEC) has shed light on the unique quantum properties of superfluids~\cite{henn_emergence_2009,neely_characteristics_2013,navon_emergence_2016,gauthier_giant_2019}, and uncovered universal aspects of fluid dynamics~\cite{reeves_identifying_2015,garcia-orozco_universal_2022,
dogra_universal_2023,madeira_universal_2024}. Quantum fluids exhibit superfluidity and support quantum vortices~\cite{onsager_statistical_1949}, and compressible Bogoliubov phonons~\cite{bogoliubov_theory_1947}, while constituent atoms can also possess internal spin degrees of freedom~\cite{tsubota_spin_2014}, or interact via long-range dipolar potentials~\cite{chomaz_dipolar_2022}. These additional degrees of freedom offer more channels for energy transport, and can cause dramatic departure from classical behavior~\cite{navon_emergence_2016,navon_synthetic_2019,gauthier_giant_2019,johnstone_evolution_2019,dogra_universal_2023}. 

A cascade of energy between scales can emerge when energy is injected into a fluid at a particular length scale and dissipated at another scale. This results in the formation of ever smaller structures down to the dissipation scale (direct cascade)~\cite{kolmogorov_local_1941}, or energy accumulating into coherent structures at the system scale (inverse cascade)~\cite{kraichnan_inertial_1967}. \revise{Where energy is only transferred between wavenumbers, $k$, of similar magnitude, the cascade is effectively local in momentum-space and power laws can develop in the energy distribution. The slope of power laws can provide clear signatures of particular dynamics, which may otherwise prove opaque.} The most well-known example of emergent power-law behavior in fluid turbulence is the Kolmogorov-Obhukov $-5/3$ law \cite{kolmogorov_local_1941}. This states that at high Reynolds number in classical incompressible fluids, the distribution of kinetic energy in Fourier-space is given by
$E(k) = C \varepsilon^{2/3} k^{-5/3}$, 
where $C$ is a universal dimensionless constant and $\varepsilon$ is the rate of energy flux per unit mass. Recent experiments \cite{navon_emergence_2016,dogra_universal_2023,navon_synthetic_2019} and numerical investigations \cite{shukla_nonequilibrium_2022,zhu_direct_2023} of turbulent BECs have reported power-laws related to predictions from weak-wave turbulence (WWT) theory~\cite{zakharov_kolmogorov_1992-1,nazarenko_wave_2011}, observed in the wave-action spectrum accessible in experiments. 

Wave-turbulence involves nonlinearly interacting waves, rather than vortices or eddies in typical hydrodynamic turbulence.  In particular, ideal WWT theory predicts a direct energy cascade with wave-action spectrum power-law $n(k) \propto k^{-3}$~\cite{dyachenko_optical_1992}. Interestingly, most of the publications report a slightly steeper scaling law than $k^{-3}$ \cite{navon_synthetic_2019,navon_emergence_2016,dogra_universal_2023}, conjectured to be due to the residual role of vortices (which are neglected in WWT theory), nonperturbative nonlinear interactions between waves, or quantum pressure~\cite{navon_emergence_2016}. Recently it has been argued that the direct energy-cascade is marginally non-local \cite{zhu_direct_2023} and that a logarithmic correction can account for the steeper power-law observed for weak forcing~\cite{navon_emergence_2016,martirosyan_equation_2024}. However, that analysis was applied to a four-wave kinetic equation, relevant for Bose gas dynamics when there is no condensate. The need for a deeper understanding of turbulent regimes in a quantum fluid motivates further analysis of the highly condensed regime~\cite{navon_emergence_2016}.

In this work, we use the Gross-Pitaevskii Equation (GPE) to simulate the development of steady turbulence in a cuboid box-trap. Energy is injected via an oscillating potential that shakes the box along one axis, and damped by selective removal of high energy atoms~\cite{navon_emergence_2016,navon_synthetic_2019}. By finely varying the forcing energy from weak to strong relative to the superfluid chemical potential we identify regimes of turbulence by their power laws, vortex distributions, and two point correlations. Motivated by a recent reformulation of spectral analysis for quantum fluids \cite{bradley_spectral_2022}, we apply those methods to the turbulent states. All quantum-phase information missing from semi-classical velocity power spectra is retained, allowing highly resolved decomposition of wave-action spectra. 

For weak forcing, we observe $k^{-7/2}$ scaling as reported in experiments and other numerical studies. \revise{Further decomposition shows that the quantum pressure is significant throughout the power-law range of $k$, consistent with the role of quantum pressure in short wavelength Bogoliubov quasiparticle excitations~\cite{bogoliubov_theory_1947,reeves_quantum_2017}}. For strong forcing, the system contains many vortices, and approaches a $k^{-7/3}$ power law for the compressible kinetic energy only, consistent with the WWT prediction of inverse particle cascade scaling in the non-condensed regime~\cite{zhu_direct_2023}. The phase coherence length of the GPE wave function drops to the healing length in this regime associated with dense vortex tangles.  

This paper is organized as follows: In \sref{secII}, we present our theoretical model of forced and damped Gross-Pitaevskii turbulence. \sref{secIII} details the identification of two distinct turbulent regimes through analysis of density fluctuations, vortex distributionwes, and the components of kinetic energy. In \sref{secIV}, we apply a reformulated spectral analysis to these regimes. Finally, in \sref{secV} we discusses the implications of our findings and present our conclusions.

\section{Theoretical Model}
\label{secII} 
Far below the critical temperature for condensation, $T \ll T_c$, Bose-Einstein condensed dilute gasses are well described by the Gross-Pitaevskii Equation~\cite{dalfovo_theory_1999}. Generalizations of the GPE to finite temperature have also been developed via several different approaches, including exact phase space methods~\cite{steel_dynamical_1998}, number-conserving Bogoliubov methods~\cite{gardiner_number-conserving_2007}, the Zaremba-Nikuni-Griffin (ZNG) formulation of two-fluid theory~\cite{zaremba_dynamics_1999}, and the truncated-Wigner classical field method~\cite{sinatra_classical-field_2001,graham_statistical_1973,polkovnikov_phase_2010}; see \cite{blakie_dynamics_2008,proukakis_finite-temperature_2008} for reviews of generalized GPE approaches to the dynamics of the dilute Bose gas.

Under conditions of weak forcing, the GPE provides a quantitative treatment of the system, directly comparable with experiments. Stronger forcing involves large energy injection and requires careful consideration. Due to the numerical scale of the system and the challenges of running large-scale simulations for very long \revise{physical times}~\footnote{Our physical evolution times of order 10s correspond to around $1300\hbar/\mu$, where $\hbar/\mu$ is the natural unit of time of the homogeneous gas.}, in this work we take the approach of constructing a minimal GPE model, similar to other recent numerical treatments~\cite{navon_emergence_2016,martirosyan_equation_2024}. Our primary differences are using higher working precision which allows consideration of stronger forcing, and larger spatial sizes in units of the healing length. Ideally, we aim to observe a power-law over a decade of wavenumbers in order to rule out the possibility of falsely attributing power-law behavior to a crossover~\cite{clauset_power-law_2009}. 

The GPE is 
\begin{align} 
    i\hbar\frac{\partial \psi (\textbf{r},t)}{\partial t}& = \left[ -\frac{\hbar^2}{2m}\nabla^2 + V(\textbf{r},t) + g|\psi(\textbf{r},t)|^2 \right] \psi(\textbf{r},t),
    \label{GPEdim}
\end{align}
where $g=4\pi\hbar^2 a/m$, $\hbar$ is the reduced Planck constant, $a$ is the $S$-wave scattering length, and $m$ is the atomic mass. \revise{For a homogeneous ground state with chemical potential $\mu$ and density $n_0 = |\psi|^2$}, we define the healing length $\xi$ via
\begin{align}
    \mu&= g n_0 =\frac{\hbar^2}{m\xi^2}
\end{align}
The natural units of time and length are $\tau=\hbar/\mu$ and $\xi=\hbar/\sqrt{m\mu}$ respectively. To make our results more concrete, we report our results using parameters for $^{87}$Rb:  $m=1.44\times 10^{-25}$kg and $a=5.8\times 10^{-9}$m. 

In Eq.~\eref{GPEdim}, the potential $V(\textbf{r},t)$, consists of a hard-walled cuboid potential, as well as a time-dependent forcing potential:
\begin{align}
    V(\textbf{r},t) &= V_\textrm{static}(\textbf{r}) + V_F(\textbf{r},t)
\end{align}
The walls of the static trapping potential are slightly smoothed rather than being piecewise defined, giving both a better description of experimental reality, and improved numerical stability due to eliminating Gibbs phenomena~\cite{piotrowska_spectral_2019}. The cuboid box trap has a large flat region with dimensions $(L_x,L_y,L_z)=(40,30,20)\xi$.  The trap walls are modeled by shifting a steep transition potential
\begin{align}
    V_\textrm{wall}(x)& = \begin{cases}
        V_b \sin(\pi x/2\ell_b)^{24} \quad\textrm{for }|x|\leq\ell_b/2\\
        V_bH(x)\quad\textrm{otherwise}
    \end{cases}
\end{align}
to coordinates $x-L_x$, etc, where $H(x)$ is the Heaviside step. Over a region of width $\ell_b=6\xi$ centered on $x=3\xi$ the  potential $V_\textrm{wall}(x)$ ramps from zero up to $V_b = 30\mu$, providing sufficiently steep walls for box-confined dynamics while avoiding overly sharp changes in the potential; the shape of the potential causes the flat density to drop away steeply at $x=0$, so that after the shift, the homogeneous region is constrained to $\pm L_x$. Numerically finding the ground state $\psi_0(\mathbf{r})$ of $V_\textrm{static}(\mathbf{r})$ using imaginary time evolution for chemical potential $\mu = k_B\times 1$nK, results in a total atom number $N \simeq 7.2 \times 10^5$. The loss of initial anisotropy of the ground state under time evolution provides a useful indicator of the development of steady isotropic turbulence.

Generation of steady-state turbulence requires both energy injection and  dissipation. Energy injection is provided by an oscillating linear potential applied along the $z$-direction in the form
\begin{align}
    V_{F}(\textbf{r},t) &= U_F\frac{z}{L_z} \sin(\omega_{F}t),
\end{align}
where $\omega_{F}$ is the driving frequency. $U_F$ is the driving energy scale, here characterising the extra energy per particle introduced at the edge of the box. To explore different turbulent regimes, we use a fixed driving frequency of $\omega_F = 4$~Hz and vary the forcing amplitude over the interval $0.1\leq U_F/\mu\leq 5$. The lower limit is a very weak perturbation, while the upper limit represents strong forcing. 

Dissipation occurs in the form of high-energy atoms escaping the trapping potential \cite{navon_synthetic_2019}. We model this numerically by introducing a dissipative `potential' outside of the box trap. The total static complex potential is    
\begin{align}
    V_\textrm{static}(\textbf{r}) &\equiv V_\textrm{box}(\textbf{r}) - iV_\textrm{diss}(\textbf{r})
    \label{Vstat}
\end{align}
where 
\begin{align}
    V_{\textrm{diss}}(\textbf{r}) &=  
    \begin{cases} 
        2.5\mu & \textrm{if } |u| > 0.5L_u + 5\xi, \hspace{2mm} u \in \{x,y,z\} \\
                0 & \textrm{otherwise.}
    \end{cases} 
\end{align}
\revise{This choice ensures that dissipation is only active outside the homogeneous region, $|u|<0.5L_u$, and introduces a dissipation scale associated with the escape energy~\cite{navon_synthetic_2019}}.
The momentum required for a particle to escape is a function of the height of the trap walls, and for our trapping potential loss occurs at dimensionless dissipation wavenumber $k_D\xi = \sqrt{2V_b/\mu} = \sqrt{60} \simeq 7.74$~\footnote{This choice also gives sufficient margin between the dissipation $k$-value and the Nyquist $k$-values for a grid of $256^3$ points (Appendix~\ref{CONM}). High-energy particles are thus dissipated well before their spatial grid representation becomes inaccurate. }. This energy and spatially selective loss is the only dissipation used in our simulations, and provides the energy sink required for a steady state to form under conditions of steady forcing.

To create turbulence in the forced and damped GPE, we apply $V_F$ to the ground state $\psi_0$ and evolve until $t=t_f\equiv 10$ seconds. This time interval ensures a steady state is reached for all $U_F$ values. We use a grid of $256^3$ points and time-evolve using a pseudo-spectral Runge-Kutta method, referred to as the RK4IP \cite{caradoc-davies_three-dimensional_2000}. A fixed time step $\Delta t=10^{-3}$s is chosen to ensure numerical convergence \revise{of} all simulations. The code was written in Julia and optimized to achieve a minimal memory footprint and run on GPU hardware using the CUDA.jl library~\cite{besard_rapid_2019,besard_effective_2019}. We achieved a runtime of 38 minutes per second of dynamics on NVIDIA A100 GPU using Float64 representation. For details of the algorithm and error analysis see Appendix \ref{CONM}.
\begin{figure}[t!]
    \centering 
    \includegraphics[width=\columnwidth]{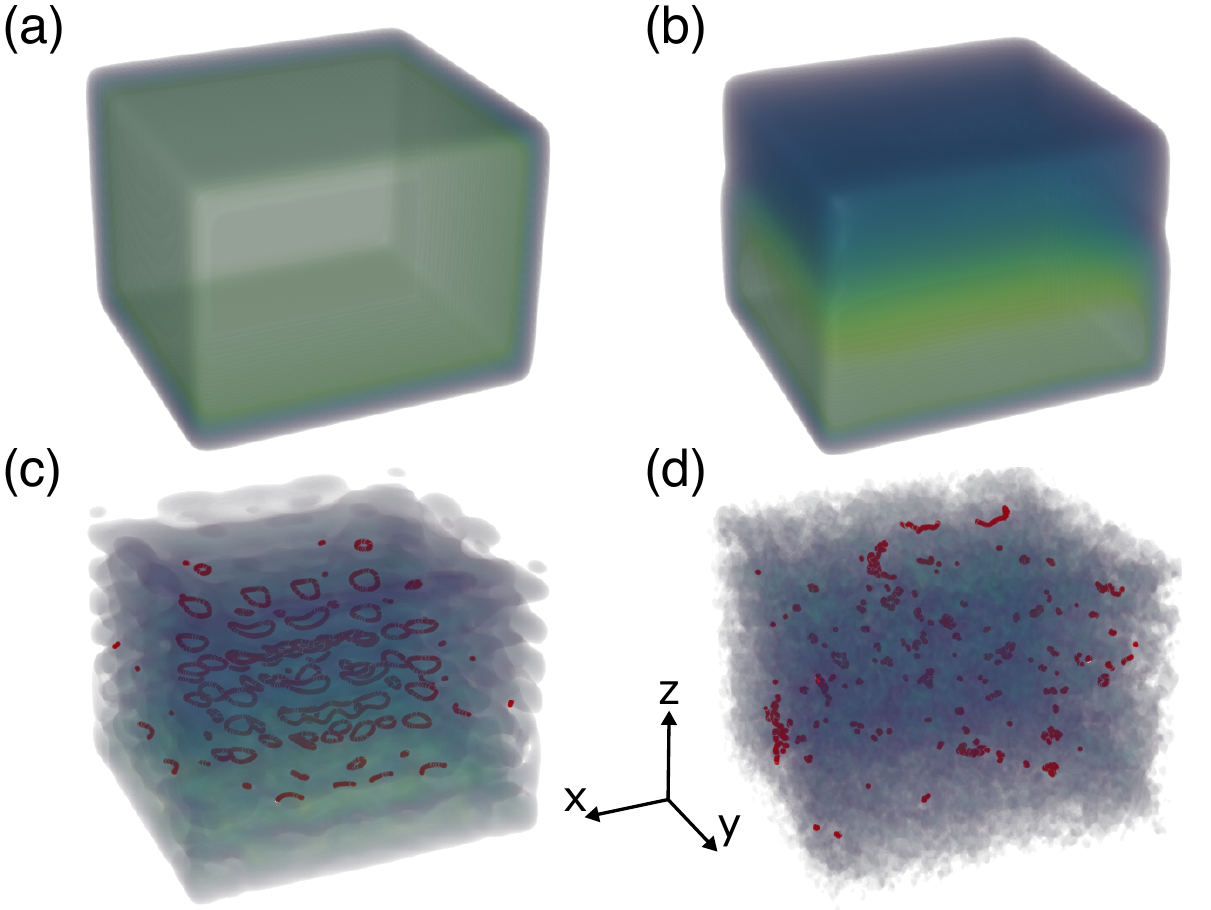}
    \caption{Time \revise{evolution} of density isosurfaces of the cuboid box trap shaken along $z$ with forcing amplitude $U_F=\mu$: (a) $t = 0.0$s shows the initial ground state of the box trap; (b) $t = 0.156$s weak density excitations along $z$; (c) $t = 0.391$s nucleation of vortex rings; (d) $t = 10.0$s steady-state turbulence. Vortex lines can be seen in red. \revise{In the steady state there are many small vortex rings on the surface, while extended long vortex lines are absent from the bulk.}}
    \label{3d_turb}
\end{figure}
\section{Development of turbulence}\label{secIII}
\subsection{Vortices and density waves}
To outline the general phenomenology, we first present the dynamics that occur at a forcing amplitude of $U_F=\mu$. The initial ground state, shown in \fref{3d_turb}(a), responds to forcing by weakly oscillating between the sides of the trap, creating weak density waves, as seen in \fref{3d_turb}(b). After sufficient energy is injected, the superfluid develops many small vortex rings parallel to the $x-y$ plane [\fref{3d_turb}(c)]. \revise{After approximately two seconds, the system transitions to a steady state, where the energy per particle oscillates about a steady mean value. The density and vortex distribution becomes isotropic, involving many vortices and significant energy, density, and phase fluctuations [\fref{3d_turb}(d)]. For this particular forcing amplitude there are many small vortex rings, but no extended vortex lines.}
        
\begin{figure*}[t!]
            \centering
            \includegraphics[width=\linewidth]{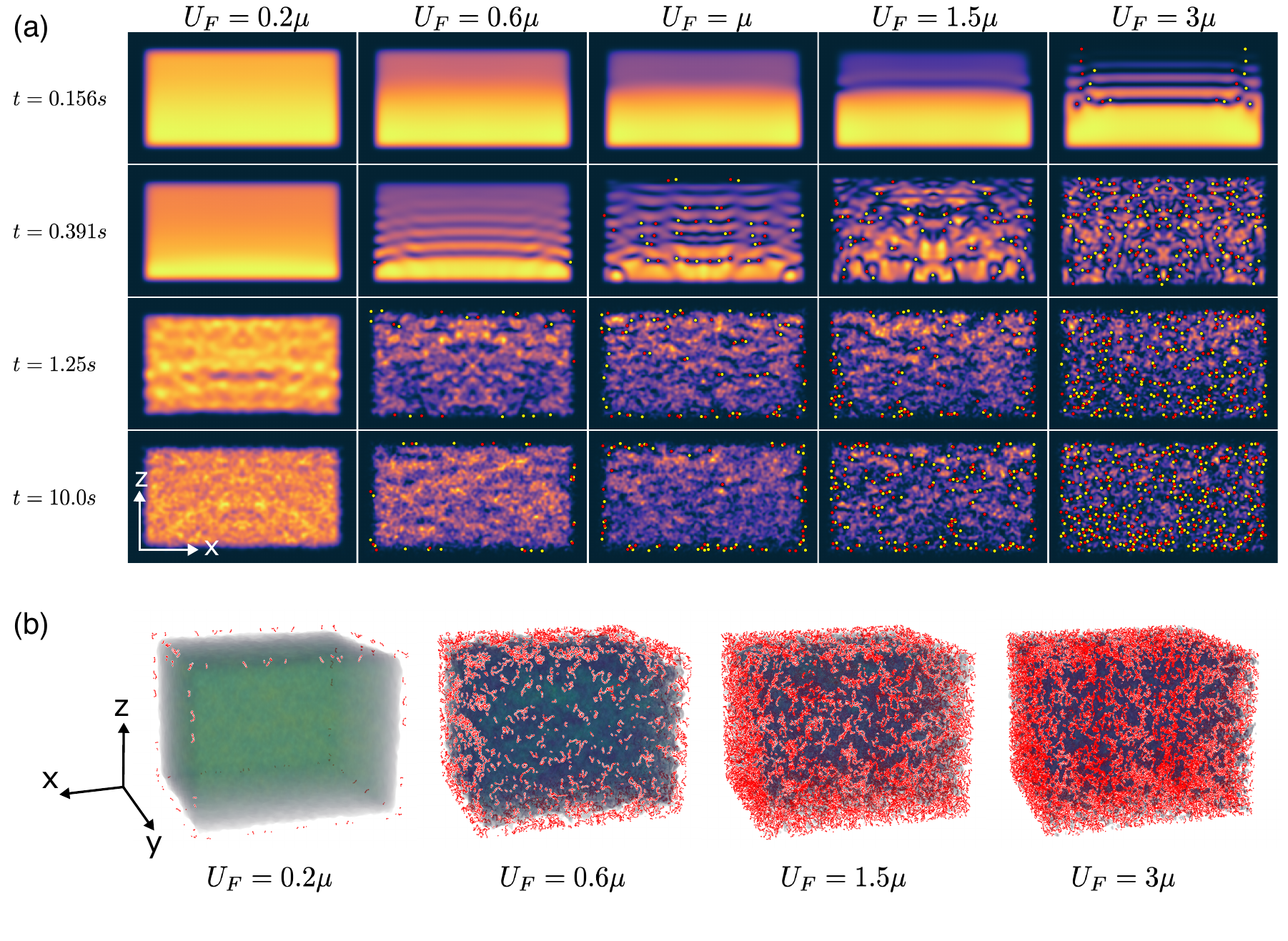}
            \caption{Development of turbulence from continuous forcing. (a) Heatmaps of slice density $|\psi(x,0,z,t)|^2$ (arbitrary units), for different forcing amplitudes, $U_F$, and evolution times $t$. Vortices and antivortices are indicated in red and yellow respectively. Isotropy develops over time from anisotropic forcing for each $U_F$, and vortices enter the bulk superfluid as $U_F$ increases. (b) Three-dimensional isosurfaces of condensate density at $t=10.0$s, corresponding to the quasi-stationary states shown in the last row of (a). As $U_F$ increases, the steady state vortex distribution (marked in red, sign omitted for clarity) undergoes a transition from dilute surface vortices ($U_F\lesssim \mu $) to dense bulk vortices ($U_F\gtrsim 2\mu $).}
            \label{heatmaps}
\end{figure*}

To study the response to anisotropic forcing along $z$, we consider condensate density and vortex distributions on the $y=0$ plane for a range of forcing amplitudes and times, shown in \fref{heatmaps}. Vortices are detected by calculating the circulation of the quantum phase around each grid plaquette in the plane. We see that weak forcing excites small amplitude fluctuations, but contains insufficient energy to generate bulk vortices, as seen in \fref{heatmaps} for $U_F=0.2\mu$. As $U_F$ increases, vortices appear at the edges of the condensate and eventually in the bulk as small rings. They do not survive as extended line vortices in the steady state until the forcing reaches $U_F\approx 1.2\mu$. For strong forcing, e.g. \fref{heatmaps}, $U_F=3\mu$, the steady state involves a dense and isotropic vortex distribution. The transition from weak surface vortices to dense bulk vortices in the steady state is evident in the three-dimensional vortex distribution, \fref{heatmaps}(b), but harder to identify due to the proliferation of surface vortices. 

\fref{excite_energies}(a) and (b) show the normalization and energy per particle of the system during excitation. We observe a brief delay between when forcing begins and when particles have sufficient energy to escape the trap, shown in \fref{excite_energies}(a). Consequently, the energy per particle of the system rapidly increases [\fref{excite_energies}(b)]. Once dissipation takes effect, the energy per particle approaches an approximate steady-state, oscillating about a mean value that increases with $U_F$; the system thus approaches an energetic steady state while continuing to lose particles.  \revise{As in \cite{navon_synthetic_2019}, atom loss is approximately linear for weak forcing. In the strongest forcing cases the rate of particle loss decreases over time, resulting in energy per particle gradually increasing.}
\subsection{Energy decomposition}
To further investigate the role of vortices and compressible excitations, we perform a decomposition of the kinetic energy~\cite{nore_kolmogorov_1997}. Using a Madelung-transform, we can represent the complex wave function (away from vortex cores) in terms of two real fields: the density, $\rho = |\psi|^2$, and the quantum phase, $\Theta$.
\begin{align}
    \psi(\textbf{r},t) &= \sqrt{\rho(\textbf{r},t)}e^{i\Theta(\textbf{r},t)}
\end{align}
Using the velocity field $\textbf{v}(\textbf{r},t) = \hbar/m \nabla \Theta (\textbf{r},t)$, we construct the density-weighted velocity field~\footnote{The ordinary velocity field  $\mathbf{v}$ diverges near the core of vortices, and so cannot be used for decomposition. By using $\mathbf{u}$ we can apply the Helmholtz decomposition to a twice continuously differentiable vector field, as required by the decomposition theorem.}, $\textbf{u} = \sqrt{\rho} \textbf{v}$, and perform a Helmholtz decomposition of the vector field into incompressible and compressible components
\begin{align}
    \textbf{u} &= \textbf{u}^i + \textbf{u}^c,
\end{align}
where the incompressible component is divergence-free and associated with vortices, and the compressible component is curl-free, associated with density excitations:
\begin{align}
    \nabla \cdot \textbf{u}^i &= 0, \; \; \; \nabla \times \textbf{u}^c = \textbf{0}.
\end{align}
We also define a quantum pressure vector field~\cite{bradley_spectral_2022}
\begin{align}
\textbf{u}^q &= (\hbar/m) \nabla \sqrt{\rho}.    
\end{align}
The quantum pressure is not a physical velocity field, but instead arises from density gradients. The total kinetic energy can then be written as the sum of the three components
\begin{align}
E_\textrm{kin}&\equiv\int d^3\mathbf{r}\frac{\hbar^2|\nabla\psi|^2}{2m}=E_\textrm{kin}^c+E_\textrm{kin}^i+E_\textrm{kin}^q,
\end{align}
where 
\begin{align}
    E_\text{kin}&=\frac{m}{2}\int d^3\mathbf{r}|\mathbf{u}^{\alpha}(\mathbf{r})|^2,
\end{align}
\revise{and $\alpha \in \{i,c,q\}$.}
\begin{figure}[t!]
    \centering
    \includegraphics[width=\columnwidth]{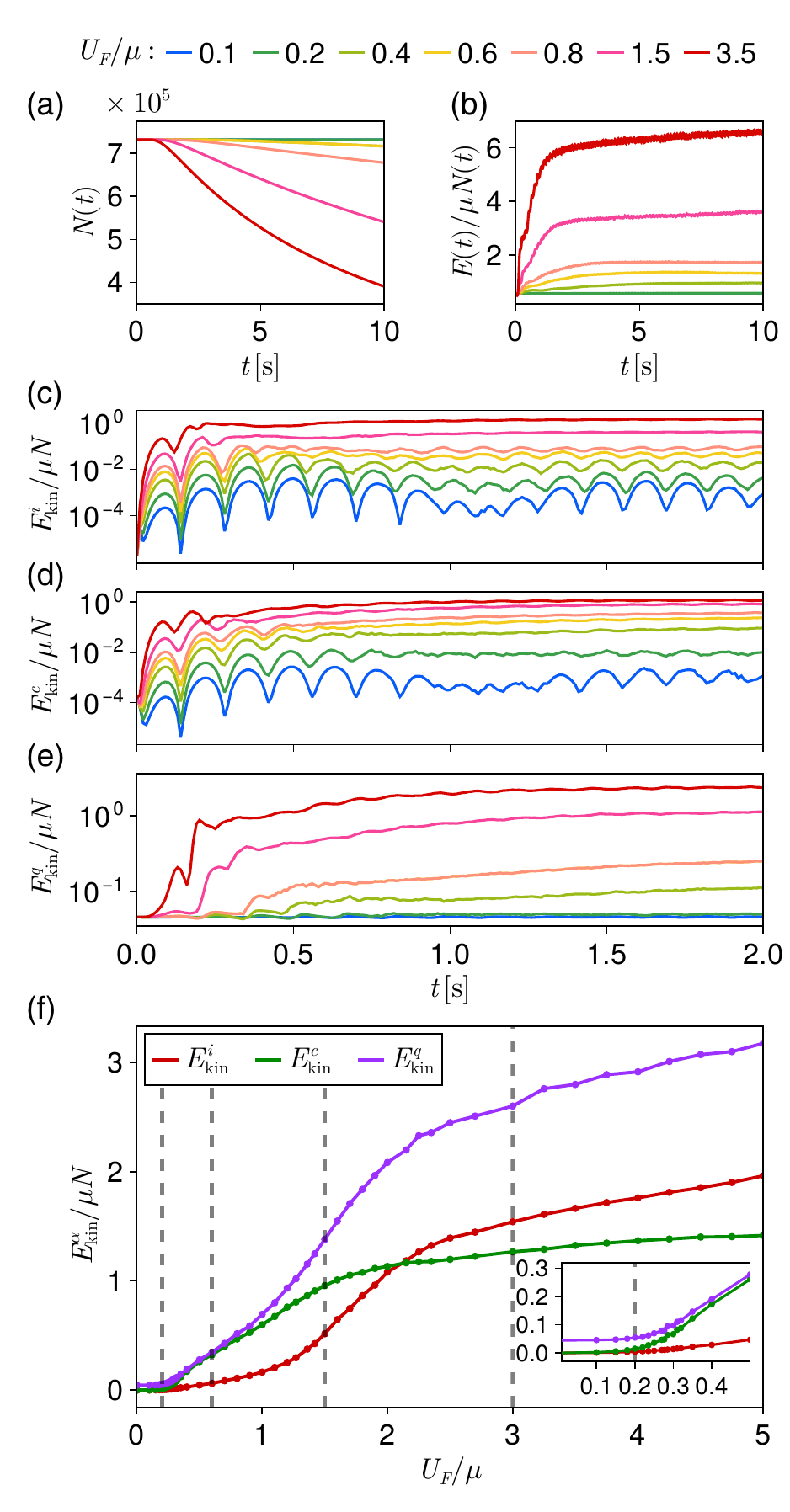}
    \caption{Development of turbulence (a) Atom number over time. The rate of particles escaping the trap increases with $U_F$. (b) Total energy per particle over time. (c), (d)\revise{, (e) Total incompressible, compressible, and quantum pressure kinetic energy during initial period of excitation.} (f) Kinetic energy fractions in the steady state ($t=10.0$s) as a function of $U_F$. Dashed lines show the four $U_F$ values of \fref{heatmaps}. }
    \label{excite_energies}
\end{figure}

Figures \ref{excite_energies}(c) and (d) show $E_\textrm{kin}^i$ and $E_\textrm{kin}^c$ during the initial period of excitation. Both components oscillate in response to the forcing, acquiring a mean value that increases with $U_F$. The kinetic energy components of the steady-state display a nonlinear dependence on $U_F$, indicating different regimes of turbulence, shown in \fref{excite_energies}(e). The relative importance of compressible and incompressible energy indicates a change from wave to vortex turbulence. \revise{Even in the absence of vortices, kinetic energy is stored in density gradients, such as those at the condensate boundary. This accounts for the nonzero $E^q_{kin}$ in \fref{excite_energies} for $U_F \sim 0$
.} Figure \ref{excite_energies}(e) is consistent with the behavior shown in \fref{heatmaps}: from $U_F = 0.2\mu$, $E^c_\textrm{kin}$ sharply increases as density fluctuations grow [\fref{excite_energies}(e)]; for the same $U_F$ values, $E^i_\textrm{kin}$ increases gradually as vortices are present only at the condensate surface. Recent reports of wave-turbulence in BECs generally correspond to these forcing amplitudes ($U_F < \mu$) \cite{navon_emergence_2016,navon_synthetic_2019,zhu_direct_2023,dogra_universal_2023}, which is consistent with the dominance of compressible kinetic energy in this region. 

At $U_F \simeq1.2\mu$, a distinct change is evident in \fref{excite_energies}(e). $E^i_\textrm{kin}$ increases rapidly, associated with the formation of bulk vortices, and $E^c_\textrm{kin}$ has a much weaker response to increased forcing. Finally, another change in response occurs for $U_F > 2\mu$, where incompressible kinetic energy and quantum pressure have a weaker response to increased forcing. In \fref{heatmaps} this corresponds to the steady state vortex distribution becoming homogeneous. There may be a limit to the number of vortices that can stably exist within a BEC of a certain size and density, bounding growth of incompressible kinetic energy with forcing. \revise{Assuming densely packed vortex rings, the bulk volume divided by the volume of a vortex ring of radius $ \sim \pi\xi$, and thickness $\sim 2\xi$, gives an estimated upper bound of $L_xL_yL_z/(2\pi^3\xi^3)\sim 400$ for our system. }

This analysis suggests distinct turbulent regimes exist for different forcing strengths. In the following sections, we apply tools of spectral analysis to relate these regimes to previous observations and theoretical predictions, aiming to identify power-law behavior associated with particular regimes in \fref{excite_energies}(e).
\subsection{Spectral Analysis}\label{speca}
    Our general aim is to identify power-law behavior in kinetic energy spectra, to gain insight into the underlying mechanism of transport and the state of the turbulent Gross-Pitaevskii field. In this section, we summarize relevant results of the spectral analysis formulation presented in Ref.~\cite{bradley_spectral_2022}, that will allow such identification.
    
For any two complex vector fields, $\textbf{u}$ and $\textbf{v}$, the spectral density of their inner product $\langle\mathbf{u}\|\mathbf{v} \rangle(k)$, is defined as 
\begin{equation}
    \langle \textbf{u} | \textbf{v} \rangle \equiv \int_0^\infty dk \langle\textbf{u} \|\textbf{v} \rangle(k).
    \label{spec_dens}
\end{equation}
It can be shown~\cite{bradley_spectral_2022} that in three dimensions 
\begin{align}
    \langle\textbf{u} || \textbf{v} \rangle(k) &= \frac{k^2}{2\pi^2} \int d^3\textbf{r} \; \frac{\sin{(k |\mathbf{r}|)}}{k|\mathbf{r}|} C\left[\textbf{u},\textbf{v} \right](\textbf{r}),
    \label{spec_IP}
\end{align}
where $C\left[\textbf{u},\textbf{v} \right](\textbf{r})$ is the spatial two-point correlation of $\textbf{u}$ and $\textbf{v}$:
\begin{equation}
    C\left[\textbf{u},\textbf{v} \right](\textbf{r}) \equiv \int d^3 \textbf{R}\; \langle\textbf{u}| \textbf{R} - \textbf{r}/2 \rangle\langle \textbf{R} + \textbf{r}/2|\textbf{v}\rangle.
\end{equation}
This reformulation of the spectral analysis problem offers several advantages. First, carrying out the angular integral in Fourier-space analytically, removes the need for numerical binning to create spectral functions of $k=|\mathbf{k}|$. Second, by decoupling the position and Fourier-space resolution, it allows arbitrary resolution spectra to be computed. Finally, a corresponding system-averaged two-point correlation can easily be extracted~\cite{bradley_spectral_2022}:
\begin{align}
    G_{u v}(r)&=\frac{1}{4\pi} \int d^3 \mathbf{r}^{\prime} \delta\left(r-\left|\mathbf{r}^{\prime}\right|\right) C[\mathbf{u}, \mathbf{v}]\left(\mathbf{r}^{\prime}\right)\nonumber\\
    &=\int_0^{\infty} dk\; \frac{\sin{(k r)}}{kr}\langle\mathbf{u} \| \mathbf{v}\rangle(k).
\end{align}
    
Depending on which fields $\textbf{u}$ and $\textbf{v}$ are inserted into \eref{spec_IP}, a variety of spectral distributions can be generated for any given wave function. For example, choosing $\textbf{u} = \textbf{v} = \psi(\textbf{r})$, we obtain
\begin{align}
    N &= \int_0^\infty dk\; \langle \psi \| \psi \rangle(k),
    \label{N_int}
\end{align}
where the momentum spectrum, describing the particle occupation of momentum-states is
\begin{align}\label{psispec}
    \langle \psi \| \psi \rangle(k)&=\frac{k^2}{2\pi^2}\int d^3\mathbf{r}\frac{\sin{(k |\mathbf{r}|)}}{kr}C[\psi,\psi](\mathbf{r}),
\end{align}
with corresponding two-point spatial correlator of the field 
\begin{align}\label{Gpp}
    G_{\psi\psi}(r)&=\int_0^{\infty} dk\; \frac{\sin{(k r)}}{kr}\langle \psi \| \psi \rangle(k).
\end{align}

Eqs.~\eref{spec_dens} and \eref{N_int} are one-dimensional integrals in $k$-space. However, spectra in WWT literature \cite{nore_kolmogorov_1997, navon_emergence_2016,navon_synthetic_2019,zhu_direct_2023}  
are usually defined such that a three-dimensional integral returns the total inner product. This introduces a factor of $4\pi k^2$ in the definition of the \emph{wave-action spectrum}
\begin{align}\label{was}
n(k) &\equiv \frac{1}{4\pi k^2}\langle \psi \| \psi \rangle (k),
\end{align}
where 
\begin{align}
N &= \int d^3\mathbf{k} \; n(k) = \int_0^\infty dk \; 4\pi k^2 n(k)\label{momentum_integral}
\end{align}
\begin{figure}[t!]
    \centering
\includegraphics[width=\columnwidth]{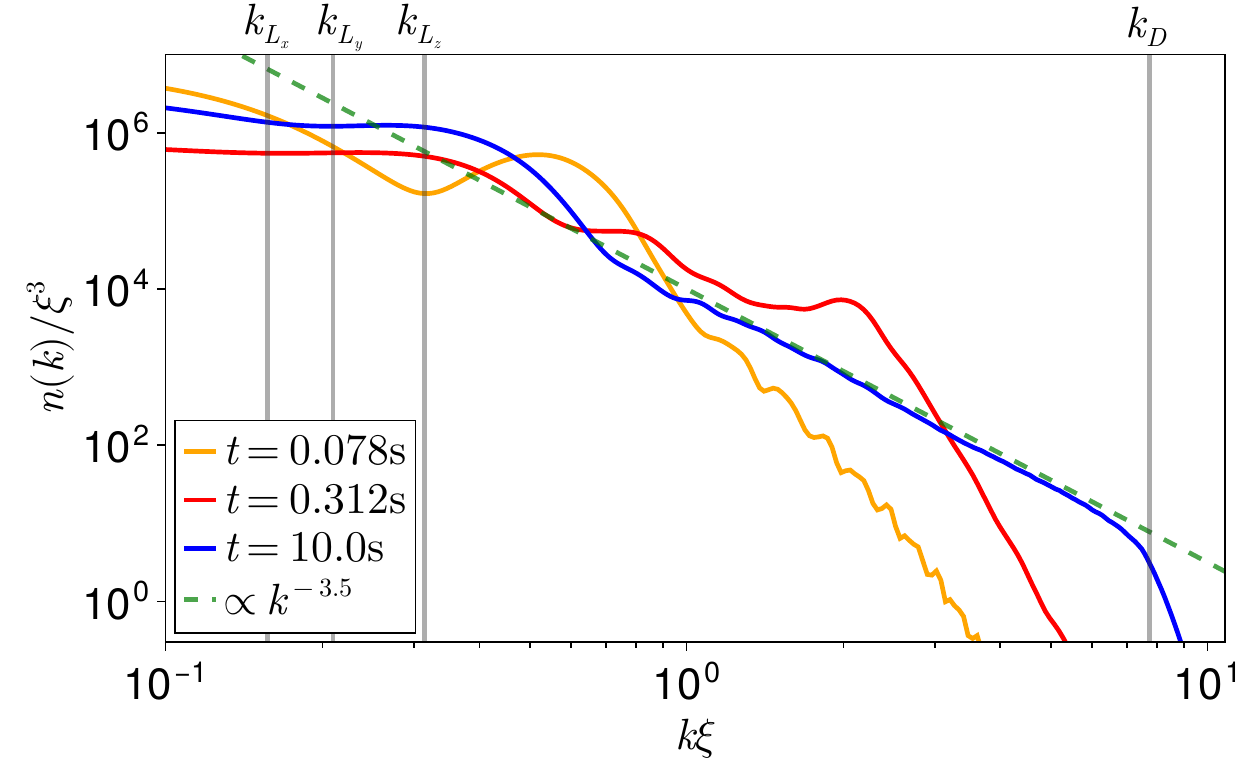}
    \caption{Evolution of the wave-action spectrum during excitation for $U_F = 0.8\mu$. Solid grey lines represent $k$-values associated with the system size $k_{L_i} = 2\pi/L_i$, and the dissipation length scale, $k_D$. As forcing is along the $z$-direction, we initially ($t=0.078$s) see a forcing peak in $n(k)$, near the length scale $k_{L_z}$. This peak then moves towards smaller lengthscales ($t=0.312$s), until reaching the dissipation length scale, $k_D$. The steady-state power law is established after $1.5$s, extending over a decade of wavenumbers until the end of the simulation ($t=10.0$s). The dashed line shows $k^{-3.5}$, and we note that the power law is clear even without time averaging.}
    \label{k3.5}
\end{figure}        
We emphasize that there is no assumption of isotropy in the definition. Instead, formally integrating over the Fourier-space solid angle generates the $\operatorname{sinc}$ kernel in \eref{psispec}. Another example is found by instead using the gradient of the wave function, $\mathbf{u}=\mathbf{v}=\nabla\psi$, giving the kinetic energy spectral-density (again introducing the spherical factor $4\pi k^2$)
\begin{equation}
    e_\textrm{kin}(k) \equiv \frac{1}{4\pi k^2} \langle \nabla \psi \| \nabla \psi\rangle(k),
    \label{ekin_definition}
\end{equation}
related to the total kinetic energy by
\begin{align}
E_\textrm{kin} &= \int d^3\textbf{k} \; e_\textrm{kin}(k) = \int_0^\infty dk \; 4\pi k^2 e_\textrm{kin}(k).
\end{align}

We now address the decomposition problem. To set up a decomposition of the kinetic energy that includes all quantum phase information~\cite{bradley_spectral_2022}, we define three fields:
\begin{align}
    (\textbf{w}^i,\textbf{w}^c, \textbf{w}^q) &\equiv (i\textbf{u}^i, i\textbf{u}^c, \textbf{u}^q) e^{i\Theta}.
    \label{w_fields}
\end{align}
These definitions furnish a linear decomposition of the gradient $\nabla \psi = (m / \hbar)[ \textbf{w}^i + \textbf{w}^c + \textbf{w}^q ] $. Using $\textbf{w}^\alpha$ as input fields to \eref{spec_dens}, we obtain the respective components of the kinetic energy-density:
\begin{align}
    e_\textrm{kin}^\alpha(k) &\equiv \frac{1}{4\pi k^2} \langle\textbf{w}^\alpha \| \textbf{w}^\alpha \rangle(k), \hspace{5mm} \alpha \in {i,c,q}.
    \label{e_kin_components}
\end{align}
The kinetic energy densities calculated in \eref{e_kin_components} retain all quantum phase information~\footnote{The standard decomposition in quantum-turbulence studies~\cite{nore_kolmogorov_1997} involves kinetic energy density calculated semi-classically as a velocity power spectrum via Plancherel's theorem (Appendix.~\ref{Parseval_deriv}), which discards quantum phase information.}. If instead we ignore the phase factor, as is usually done~\cite{nore_kolmogorov_1997}, we would obtain the standard velocity power spectra~\footnote{The usual definition does not factor out the spherical measure $4\pi k^2$, here written explicitly for convenience in connecting with recent literature.}.
\begin{align}
    \varepsilon_\textrm{kin}^\alpha(k) &\equiv  \frac{1}{4\pi k^2} 
    \langle\textbf{u}^\alpha \| \textbf{u}^\alpha \rangle(k) , \hspace{5mm} \alpha \in {i,c,q}. 
\end{align}
An important feature of the velocity power spectra components is that while they integrate to give the respective component of kinetic energy,  $E^\alpha_\textrm{kin}$, they do not sum locally in $k$-space to give the power spectral density~\cite{reeves_signatures_2014}
\begin{align}
    \varepsilon_\textrm{kin}(k) &\neq \varepsilon^i_\textrm{kin}(k) + \varepsilon^c_\textrm{kin}(k) + \varepsilon^q_\textrm{kin}(k).
\end{align}
Conversely, the components of the energy density do add locally, albeit with additional cross-terms describing redistribution of energy between scales
\begin{align}\label{allterms}   
    e_\textrm{kin}(k) &= e^{i}_\textrm{kin}(k) +    e^{c}_\textrm{kin}(k) + e^{q}_\textrm{kin}(k) \nonumber \\ 
    &+ e^{ic}_\textrm{kin}(k) + e^{iq}_\textrm{kin}(k) + e^{cq}_\textrm{kin}(k).
\end{align}

The cross-terms integrate to zero and thus do not contribute to the total kinetic energy, but are required to get the correct total energy at each $k$~\cite{bradley_spectral_2022}. In our simulations $e^\alpha_\textrm{kin}(k)$ and the corresponding $\varepsilon^\alpha_\textrm{kin}(k)$ were found to differ significantly, as detailed in Appendix~\ref{e_vs_e}.

As an example to check our formulation, we plot the time evolution of $n(k)$ in \fref{k3.5}. As forcing is initiated, a peak in the spectrum is evident at the wavenumber $k_{L_z}=2\pi/L_z$, corresponding to the forcing lengthscale. This peak then migrates towards higher $k$-values, as expected for a direct cascade. Shortly after reaching the dissipation length scale, $n(k)$ remains approximately constant in time apart from a change in amplitude due to particle loss. For this particular forcing, the power law form $n(k)\sim k^{-3.5}$ is evident in \fref{k3.5}, consistent with previous works~\cite{navon_emergence_2016,navon_synthetic_2019,martirosyan_equation_2024}. 

\section{Steady State turbulence}\label{secIV}
We now investigate the properties of steady-state turbulence. We use the spectral analysis developed in the previous section to further decompose the wave-action spectrum. As kinetic energy densities and number densities only differ by a factor of $\hbar^2k^2/2m$, the spectral decomposition~\eref{allterms} can be used to define the component wave-action spectra 
\begin{align}
    n^\alpha(k) & = \frac{2m}{\hbar^2k^2} e^\alpha_\textrm{kin}(k) = \frac{m}{2\pi \hbar^2k^4} \langle\textbf{w}^\alpha \| \textbf{w}^\alpha \rangle(k), 
    \label{nk_i}
\end{align}
where $\alpha \in \{i,c,q\}$. 
We can then use the previously established decomposition of energy-densities to identify power-law behavior in distinct parts of the wave-action spectrum, considering the compressible, incompressible, and quantum pressure components~\footnote{We do not consider the cross terms further in this work as they are not very intuitive to interpret. See Ref.~\cite{bradley_spectral_2022} for a simple example.}. 
\begin{figure}[!t]
        \centering
        \includegraphics[width=.95\columnwidth]{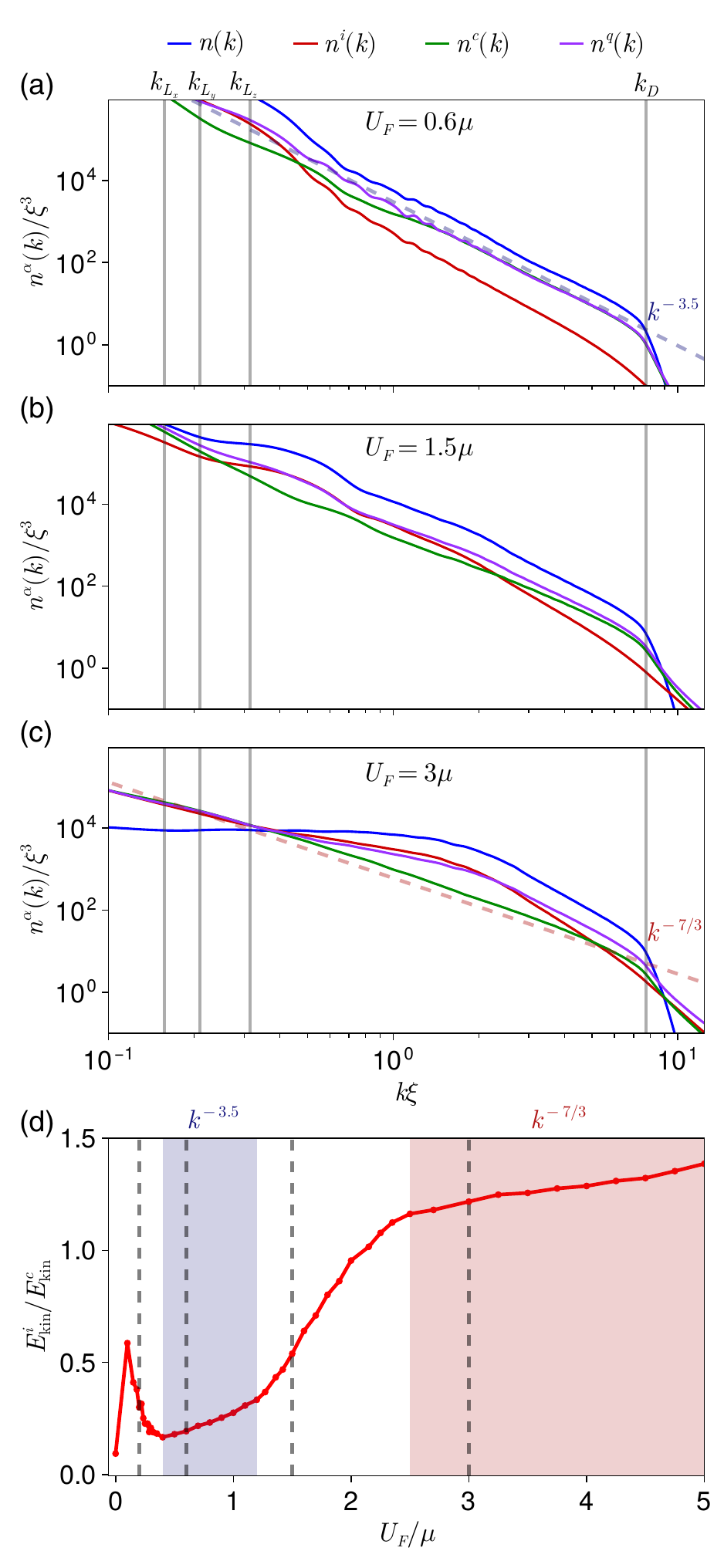}
        \caption{(a)-(c) Decomposed momentum distributions for representative forcing amplitudes of \fref{heatmaps}, averaged over one forcing period. (a) As forcing increases a $k^{-3.5}$ scaling regime is evident in both compressible and quantum pressure components. (b) Compressible and quantum pressure components decouple, associated with bulk vortices. \revise{Power-law behavior in this region extends only over a very small scale range.} (c) For very strong forcing, the compressible wave-action spectrum approaches a $k^{-7/3}$ power-law form. (d) Ratio of total incompressible to compressible kinetic energy, with identified power law regimes. $U_F$ values from \fref{heatmaps}(b) are shown as dashed lines. 
        } 
        \label{decomposed_nk}
        \vspace{2pt}
    \end{figure}
We decompose the steady-state wave-action spectrum according to \eref{nk_i} and present the densities $n^{\alpha}(k)$ in \fref{decomposed_nk}. The decomposed spectra reveal further information beyond what is available from $n(k)$. The presented steady-state spectra are averaged over one forcing period, however, time averaging only alters the result near the forcing lengthscale, a further indication that the steady-state is reached. 

We do not observe any obvious power-law behaviour for $U_F\ll \mu$ (not shown). Within a narrow region of forcing, $0.15\mu \lesssim U_F \lesssim 0.3\mu$, the compressible density approximates $k^{-3}$. However, the approximate scaling only extends over a factor of 2 in wave number, insufficient to identify power-law behaviour~\cite{clauset_power-law_2009}. 
Starting at $U_F=0.4\mu$, we observe a clear $k^{-3.5}$ scaling over nearly a decade of $k\xi$, in both compressible and quantum pressure momentum distributions, shown in \fref{decomposed_nk}(a). This power-law remains intact for a wide interval of forcing energies, $0.4\mu \leq U_F\leq 1.2\mu$, and as incompressible energy remains small, the power-law is also evident in the total spectrum $n(k)$. This is consistent with the observed vortex configurations [\fref{heatmaps}(a)], where extended bulk vortices remain rare. \revise{Despite resembling a power-law, the slope of $n^i(k)$ corresponds to a uniform distribution of incompressible energy in $k$-space.} In this regime it is notable that $n^c(k)\simeq n^q(k)$ in the power law region of wavenumbers. We return to this point in~\sref{discussion}. 
        
For $U_F \gtrsim 1.2\mu$, extended vortices develop in the bulk superfluid in the steady-state, and the system enters a transitional regime without clear power-law behavior in any component of kinetic energy [\fref{decomposed_nk}(b)]. The total wave-action spectrum gradually steepens in the ultraviolet due to an increase in the incompressible energy-density, while the compressible and quantum pressure components uncouple, and the amplitude of $n^c(k)$ gradually decreases. The appearance of bulk vortices is associated with the breakdown of $k^{-7/2}$ scaling. Eventually, we observe a wide $n^c(k) \propto k^{-7/3}$ power-law for strong forcing, $U_F\gtrsim 2\mu$, in agreement with the predicted inverse particle cascade for weak forcing in the absence of a condensate [\fref{decomposed_nk}(c)] ~\cite{zhu_direct_2023}.  

To further understand the different regimes of steady turbulence, in \fref{decomposed_nk}(d) we plot the ratio of incompressible to compressible kinetic energy, with the identified power-law regimes. There is a clear change in the energy dependence on $U_F$ at the boundary of each power-law regime. Initially, the ratio is rapidly decreasing, due to compressible excitations paired with a lack of vortex nucleation. With additional energy injected, this is succeeded by a sharp increase in the kinetic energy ratio (as surface vortices begin to proliferate), and the appearance of a $k^{-3.5}$ power-law. The breakdown of $k^{-3.5}$ involves extended vortices entering the bulk in the steady-state, and a steep rise in incompressible energy with $U_F$. The system enters a regime of vortex turbulence, but shows no scaling behavior. Finally, for $U_f\gtrsim 2\mu$, a $k^{-7/3}$ scaling emerges where vortices are approximately homogeneous and dense, and the incompressible energy responds only weakly to increased forcing. In this regime $E^i_\text{kin}/E^c_\text{kin} > 1$, a regime we can regard as strong quantum vortex turbulence, in contrast to weak wave turbulence where $E^i_\text{kin}/E^c_\text{kin} \ll 1$.

\begin{figure}[h!]
    \centering
    \includegraphics[width=\linewidth]{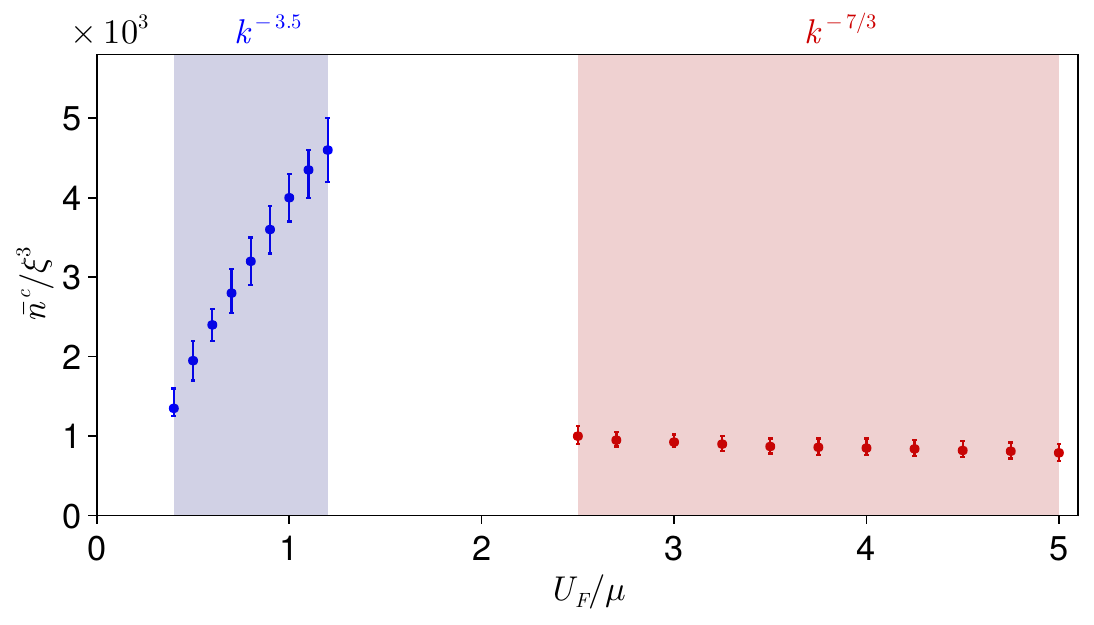}
    \caption{Power law amplitude as a function of forcing strength. $\bar{n}^c$ is shown for power laws $n^c(k) = \bar{n}^c (k\xi)^{-3.5}$ (blue), \fref{decomposed_nk}(a), and $n^c(k) = \bar{n}^c (k\xi)^{-7/3}$ (red), \fref{decomposed_nk}(c). Error bars express the range of n values for which the power law solution intercepts at least $50\%$ of points in the inertial range. Between the two regimes, the magnitude of $n^c(k)$ decays and there is no obvious power law behavior.}
    \label{amplitudes}
\end{figure}

\begin{figure}[b!]
    \centering
    \includegraphics[width=\columnwidth]{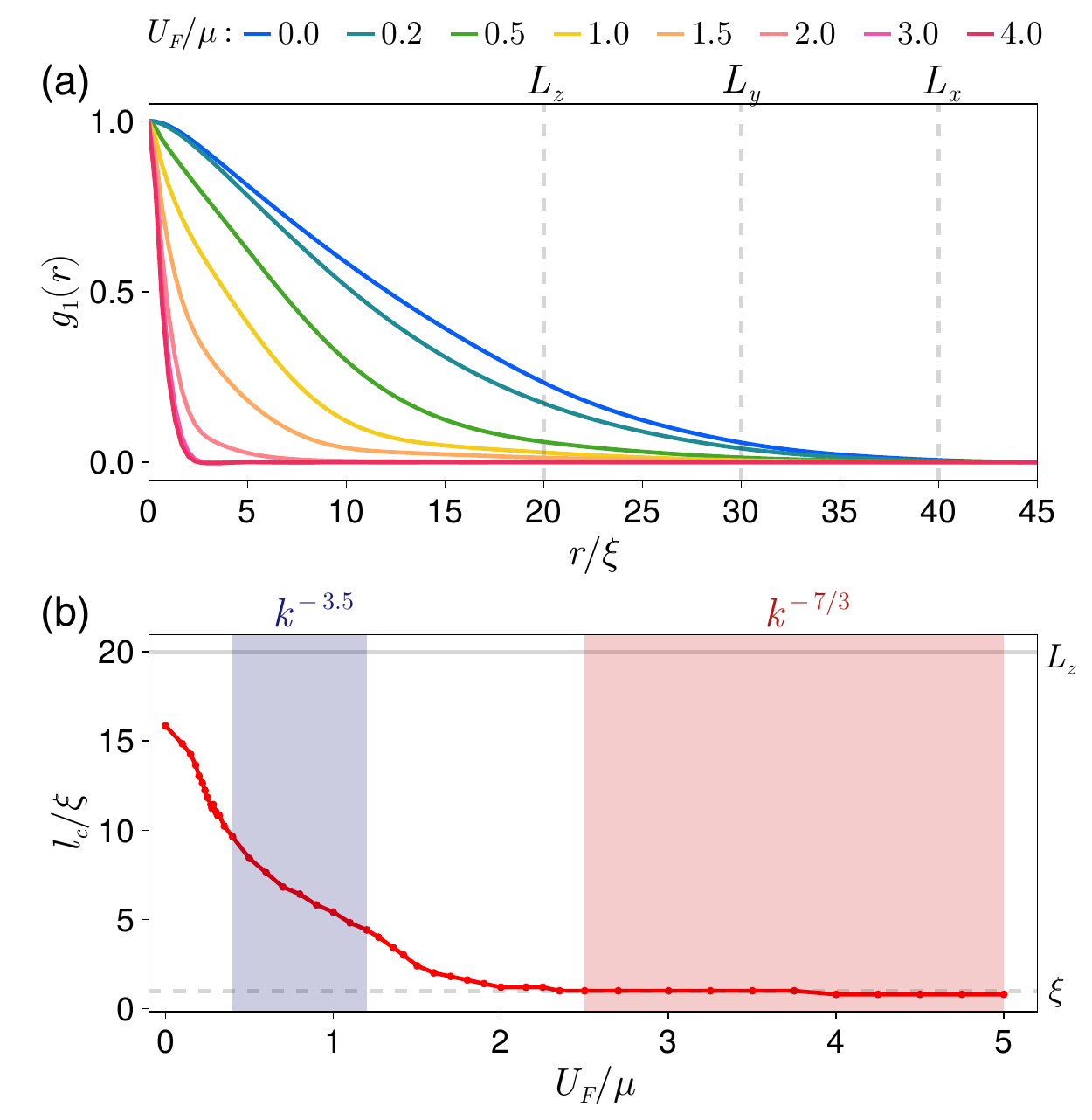}
    \caption{Correlation function for different forcing energies: (a) System averaged two-point correlation function $g_1(r)$. (b) Correlation length determined as distance where $g_1(l_c)=e^{-1}$. For weak forcing, $l_c$ is comparable to the smallest system dimension $L_z$, while for strong forcing it approaches the healing length $\xi$. The bands indicate the $-3.5$ and $-7/3$ power law regions for compressible kinetic energy.} 
    \label{two_point}
\end{figure}

\revise{For a power law of the form $n(k) = \bar{n}(k\xi)^{-\alpha}$, the amplitude of the power law, $\bar{n}$, can be extracted by plotting a compensated spectra $n(k)(k\xi)^\alpha$ (see \fref{zhu_comp} for an example of a compensated spectrum). \fref{amplitudes} shows the amplitudes for the two power laws observed in $n^c(k)$. The $k^{-7/3}$ regime shows little variation with forcing strength, reflecting the states saturation with densely packed vortices. In contrast, the $k^{-3.5}$ regime demonstrates a proportional increase in the amplitude of the spectrum with forcing.}

While Fourier space reveals power-law behavior, we can also examine the corresponding correlation functions in position-space. When power-law behavior weakens, the appearance of characteristic length scales can be more clearly understood in position space. Fourier transformation of the momentum distribution gives the system-averaged spatial two-point correlation of the Gross-Piteavskii wave function, as discussed in \sref{speca}. In \fref{two_point}(a), we plot the angle-averaged two-point correlation function $g_1(r)\equiv G_{\psi\psi}(r)/G_{\psi\psi}(0)$, calculated by transforming $\langle \psi\|\psi\rangle(k)=4\pi k^2n(k)$ to position space via \eref{Gpp}.
The two-point correlator of the ground state ($U_F=0$) decays on a scale comparable to the shortest cuboid length $L_z$. As $U_F$ increases, the length scale is reduced. Eventually, for strong forcing the correlations only persist over a scale of order $\xi$. We define the correlation length $l_c$ as the distance where $g_1(r)$ drops to $1/e$. In \fref{two_point}(b) we plot the correlation length against $U_F$, together with the power law regimes identified.

Initially there is a steep drop, followed by a slower decline during the $k^{-3.5}$ regime. As vortex lines develop in steady-state, we see a second steep decline in correlation length around $U_F\sim1.5\mu$. The dense vortex regime $U_F\gtrsim2\mu$ corresponds to $l_c\sim\xi$, and the system loses all coherence beyond the vortex core scale.   

\section{Discussion and Conclusions}\label{secV}
\subsection{Discussion}\label{discussion}
We identify three distinct regimes of steady turbulence through a combination of power laws, vortex distributions, and two point correlations.

\emph{Weak wave turbulence.---} For weaker forcing, $0.4\mu\lesssim U_F\lesssim 1.2\mu$, there is a $\propto k^{-3.5}$ power law spectrum for compressible kinetic energy, quantum pressure, and the total spectrum, and a distinct absence of extended vortex lines in the bulk superfluid. In this weak-wave turbulence regime $E^i_\text{kin}/E^c_\text{kin}\ll 1$. Even when small vortex rings appear  in the steady state as $U_F$ approaches $\mu$, the $k^{-3.5}$ scaling is robust and spans a decade of $k$-space. The forcing amplitude reported in Ref. \cite{navon_emergence_2016}, $U_F = 0.8\mu$, is in the center of the $U_F$ range where we identify the scaling both instantaneously~[\fref{k3.5}] and in time averages~[\fref{decomposed_nk}]. In addition, we observe a clear forcing peak near the system scale $L_z$ in the total wave-action spectrum at early times \revise{[Fig.~\ref{k3.5}]}. The wave-action spectra decomposition, combined with absence of bulk vortices implicates a direct Bogoliubov quasiparticle cascade for the $k^{-3.5}$ power law. Indeed, Bogoliubov phonons contain a quantum pressure component associated with small length-scale density fluctuations when $k\xi\gtrsim 1$. In \fref{decomposed_nk}(a), we see $n^c(k)$ and $n^q(c)$ are closely matched within this $k$ range, and decouple when $k\xi \lesssim 1$. \revise{The role of quantum pressure may be seen by linearizing the GPE around a homogeneous condensate and using the Madelung decomposition~\cite{reeves_quantum_2017}. For Hamiltonian GPE evolution, the superfluid Euler equation together with continuity can be used to derive the equation of motion for linearized density fluctuations $\delta n \equiv n - n_0$. They obey the generalized wave equation
\begin{align}
    \frac{\partial^2 \delta n}{\partial t^2}=\frac{gn_0}{m} \nabla^2 \delta n-\left(\frac{\hbar}{2 m}\right)^2 \nabla^4 \delta n.
\end{align}
The first term corresponds to the hydrodynamic regime, while the second term generates the free-particle behavior at high $k$ and stems from the quantum pressure term in the GPE.
Assuming plane wave solutions for the density $\delta\rho \sim \Re\{\delta\rho_0 \exp{[i(kr-\omega t)]}\}$, we recover the Bogoliubov dispersion relation
\begin{align}
    \hbar\omega = \sqrt{\frac{\hbar^2k^2}{2m}\left(2gn_0+\frac{\hbar^2k^2}{2m}\right)}.
\end{align}
A phonon cascade at high $k$ will necessarily involve significant quantum pressure, due to its role in short wavelength quasiparticle excitations. 
}

The log-corrected spectrum analysis of WWT direct cascade is derived for very weak forcing in the absence of a condensate~\cite{zhu_direct_2023}, and is thus not directly applicable to our system that involves a large condensate and hence three-wave interactions. However, the three-wave kinetic equation also supports a $k^{-3}$ power law~\cite{nazarenko_wave_2011}. \revise{Furthermore, we can expect four-wave interactions to dominate when either (i) the condensate is highly depleted due to strong forcing, or (ii)  the interaction is centered around high $k$ values where there is no condensate.} Applying the log-corrected $k^{-3}$ form to our wave-action spectra, we find good agreement provided we identify the forcing wavenumber for phonons as $k_F=1/\xi$, instead of the natural forcing wavenumber of the potential $2\pi/L_z$ as detailed in Appendix~\ref{logmodel}. Our findings are consistent with the analysis of Martirosyan \emph{et al}.~\cite{martirosyan_equation_2024}, who reached a similar conclusion~\footnote{Their result translates to $k_F=\sqrt{2}/\xi$ in our units. However, for this estimate of the forcing scale we see a narrower flat region in the log-corrected spectrum. For further discussion see Appendix~\ref{logmodel}.}.

\emph{Mixed turbulence.---} For $1.2\mu\lesssim U_F\lesssim 2.5\mu$ the system enters a transitional regime where the $k^{-3.5}$ power-law breaks down, and power law behavior is not strongly evident in any components. The compressible and quantum pressure components decouple in $k$ space, and the incompressible energy increases rapidly with $U_F$, entering a regime where $E^i_\text{kin}/E^c_\text{kin}\sim 1$, and the compressible and incompressible wave-action spectra become comparable in $k$ space. This is associated with the appearance of extended vortices in the bulk superfluid. The transition is also evident in the two-point correlation function, where the correlation length decreases rapidly as vortices enter the bulk.

\emph{Strong vortex turbulence.---} For $2.5\mu\lesssim U_F$, the condensate bulk becomes saturated with vortex lines and the compressible spectrum approximates a $k^{-7/3}$ power-law over more than a decade of $k$; this is consistent with the predicted WWT scaling for an inverse particle-cascade solution to the four-wave kinetic equation. In Ref.~\cite{zhu_direct_2023}, this was observed in the total spectrum for weak forcing in the absence of BEC, while here we see it in the compressible spectrum only, for strong forcing. As the power law extends far below $k_\xi=1/\xi$, this would be consistent with WWT if random vortices effectively destroy the condensate, and indeed the phase coherence length drops to the healing length at $U_F\gtrsim 2\mu$ [\fref{two_point}]. As our model only contains damping for high wavenumbers, a steady inverse-cascade of particles can only exist in $n^c(k)$ if particles are transported to high $k$ by the other components of $n(k)$. Physical consistency thus requires a compressible particle source near the dissipation scale $k\sim k_D$ to generate an inverse cascade; one candidate mechanism is the steady annihilation of vortices, creating bursts of sound at small scales. In Figure \ref{excite_energies}(d),(e) the compressible energy driving by the forcing appears to become ineffective for $U_F\gtrsim 2\mu$, \revise{there is also negligable variation in spectrum amplitude with forcing [\fref{amplitudes}]}, supporting the interpretation that compressible energy is instead injected at small scales via vortex annihilation.

In the high-energy regime, the incompressible energy density is extremely stable across both $U_F$ values and forcing phase. In all other cases, spectra visibly oscillate in the infrared with forcing, necessitating cycle-averaging of spectra. However, the dense distribution of vortices appears to depend very weakly on the phase of the forcing. Wave-turbulence closure requires the initial phase and amplitude of the Gross-Pitaevskii field be independent and random, whereas we find that for weak forcing there is still a high degree of coherence. Conversely, once the condensate becomes saturated with vortices, we see a drastic decrease in the correlation length, consistent with the assumptions underlying WWT. Despite strong forcing, wave interactions may be sufficiently weak that WWT theory remains relevant, particularly if injected energy primarily resides in the quantum pressure and incompressible components of kinetic energy.

Calculation of spectra with arbitrary infrared $k$-resolution and decomposition of wave-action spectra require the reformulated spectral densities~\cite{bradley_spectral_2022}. In Appendix \ref{e_vs_e} we compare the decomposition via the wave-action spectrum with the standard decomposition of the the velocity power spectrum used in previous studies. We find the two are similar only in a minority of cases, typically in the inertial range for well-developed turbulence.
\subsection{Conclusions}
Motivated by recent experiments in forced quantum turbulence~\cite{navon_emergence_2016,navon_synthetic_2019,dogra_universal_2023}, we have simulated a Gross-Pitaevskii model for a wide range of forcing energies, generating and analyzing steady turbulence. We find evidence for regimes of weak-wave and strong-vortex turbulence separated by a mixed regime that lacks clear power-law behavior. Distinct regimes of power-law behavior are identified by application of high resolution spectral analysis~\cite{bradley_spectral_2022}, and further studied through vortex distributions and two-point correlations. Our results are largely consistent with other recent works~\cite{martirosyan_equation_2024}, while also enabling further decomposition of the wave-action spectrum. For weak forcing, the bulk superfluid is approximately vortex free, and exhibits a power-law of $\sim k^{-3.5}$, consistent with weak-wave turbulence predictions for a direct cascade~\cite{zhu_direct_2023}. The compressible and quantum pressure components also show the same power-law scaling and are in close agreement. Log-correction analysis implicates a forcing wavenumber $k_F\sim \xi^{-1}$ for the cascade. In this regime the superfluid order parameter remains coherent over the system scale, with coherent only weakly degraded by small vortex rings. In contrast, for strong forcing the steady state develops a dense tangle of extended vortices that limits phase coherence to the healing length scale. Power-law scaling $\sim k^{-7/3}$ is observed in the compressible component only, consistent with the weak-wave turbulence prediction of an inverse cascade in non-condensate four-wave kinetics~\cite{zhu_direct_2023}.

Several open questions would benefit from future exploration. It would be interesting to study the rapid vortex growth regime in the transition from $k^{-3.5}$ to $k^{-7/3}$ scaling, together with the different kinetic equation approaches and their associated scaling solutions. Another open question is identifying the true source of high-$k$ compressible energy in the $k^{-7/3}$ regime. Detailed study of vortex tangles, energy fluxes, and velocity increments~\cite{zhao_kolmogorov_2024} would also reveal further insights into the nature and formation of quantum turbulence. 

\section*{Acknowledgements}
We thank Gevorg Martirosyan, \revise{Nir Navon}, and Tim Copland for stimulating discussions.
TF would like to thank the University of Otago for financial support.

The authors wish to acknowledge the use of New Zealand eScience Infrastructure (NeSI) high performance computing facilities as part of this research.

\providecommand{\noopsort}[1]{}
%

\appendix
\section{Numerical method}\label{CONM}
Evaluation of the potential term in the GPE, as well as the interaction term $|\psi(\textbf{r},t)|^2$, is straightforward using element-wise operations. However, the kinetic term contains a Laplacian, which is less straightforward. By applying a Fourier-transform to the wavefunction, we shift to momentum-space, where the kinetic operator becomes diagonal. This method is commonly known as pseudo-spectral differentiation and makes use of the Fourier-relation
\begin{align}
    \nabla^2 \psi (\textbf{r},t) \leftrightarrow -k^2 \phi(\textbf{k},t),
\end{align}
where the arrows denote a spatial Fourier-transform, and $\phi(\textbf{k},t)$ is the Fourier-transform of $\psi(\textbf{r},t)$. This method is efficient and offers excellent error properties compared to alternatives; however, it does come with some limitations. One limitation is that the boundary conditions imposed on the numerical grid become necessarily periodic. Another issue is that aliasing can occur when representing certain momentum values, based on the grid's density. Several tests were performed to ensure these factors were appropriately mitigated. 

We begin by ensuring that a sufficiently fine grid is being used to represent the system and that numerical error is within an acceptable margin. We test this only for the most violent and energetic case ($U_F = 5\mu$). The error for smaller forcing amplitudes is expected to be approximately bounded by this case. Excitation was simulated across a range of step-sizes and grid points, with no damping potential applied. In the absence of damping, normalisation of the wavefunction should be conserved, allowing any deviation to be attributed to numerical error. By monitoring the normalisation over time, we can quantify the error for each case. 
\begin{figure}[t!]
\centering
\includegraphics[width=1\linewidth]{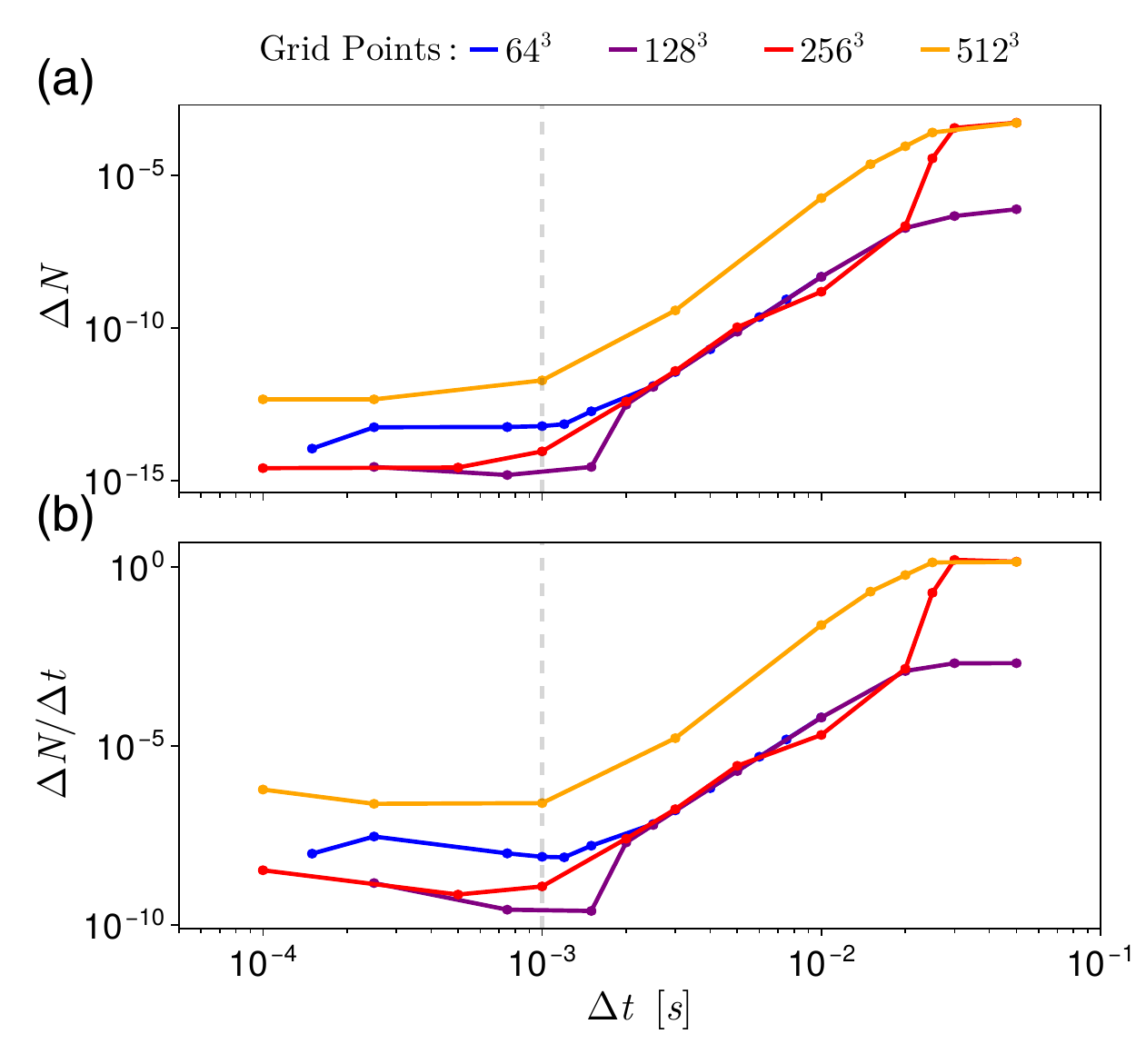}
    \caption{Error vs stepsize. (a) Average change in norm per time-step. (b) Average change in norm per second. Smaller timesteps correspond to more timesteps per second of simulations, therefore increasing error if the error per timestep does not decrease. All grid sizes  have similar error values, except $512^3$, which is greater.}
    \label{error_tests}
\end{figure}
As we would expect, \fref{error_tests}(a) shows the error per time-step decreasing with step-size, until reaching a floor determined by the limit of Float64 precision. Reducing step-size beyond this point increases the total error [\fref{error_tests}(b)], as more steps are required per second of dynamics. For this reason a step-size of $\Delta t = 10^{-3}$ was chosen for calculations. 

After selecting the step-size, we determine the necessary grid density to ensure convergence. Similar to step-size, an overly dense grid is not necessarily optimal. Although it does not increase error, it significantly raises memory-usage and computation time. The number of points along each axis of the grid, $M$, was chosen as powers of two to ensure optimal performance under the FFTW algorithm. The total number of grid points is then $M^3$. To test for convergence, we performed identical excitation of a ground state using various grid sizes. We then calculate the total kinetic energy, compressible kinetic energy, and the wave-action spectrum for each case. 
    
\begin{figure}[t!]
    \centering
    \includegraphics[width=\columnwidth]{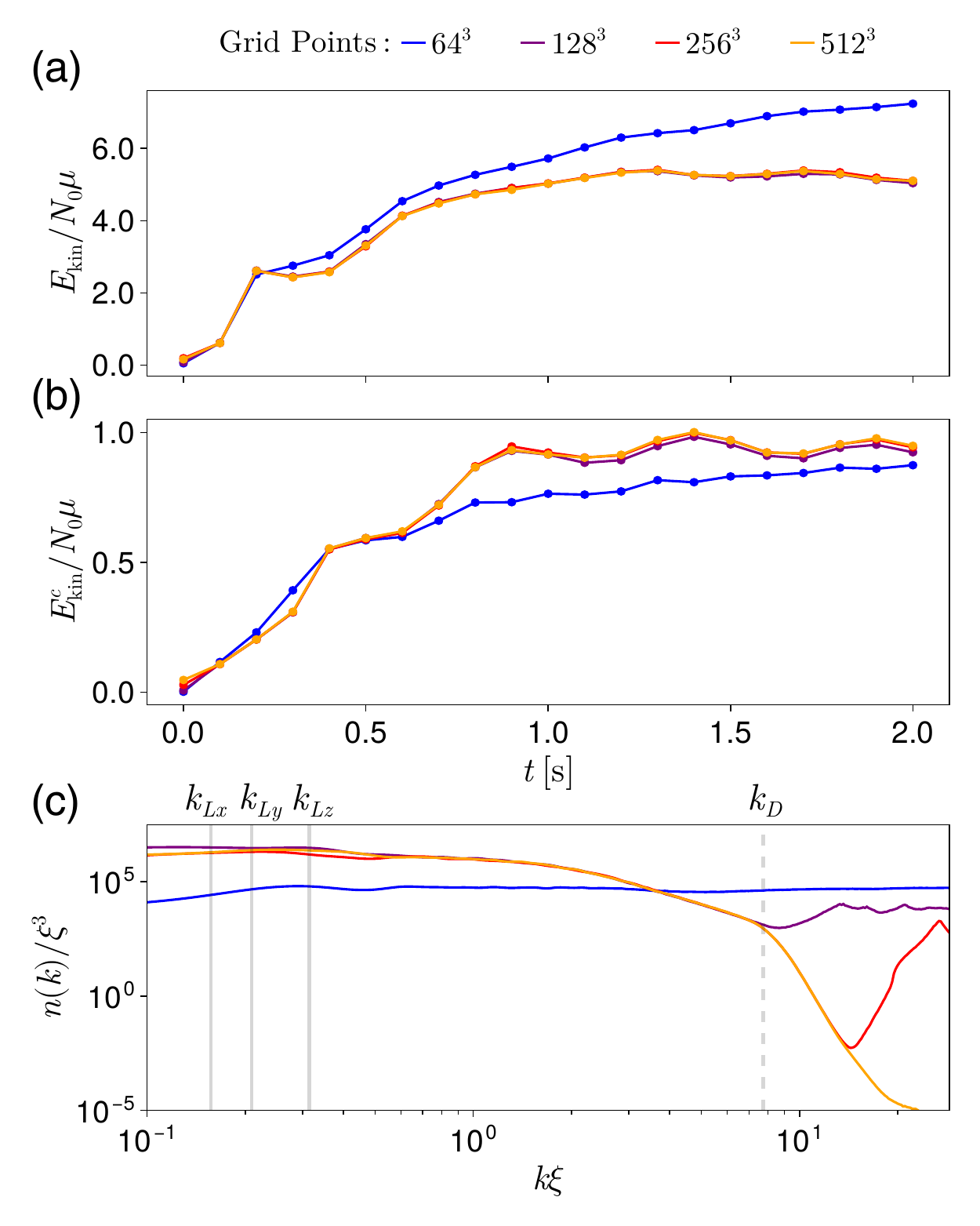}
    \caption{Convergence tests using varied grid sizes. (a) Total kinetic energy. (b) Compressible kinetic energy. (c) Wave-action spectra. (a) \& (b) show good agreement for grids larger than $64^3$. In (c) we can see that the maximum $k$-value before numerical artifacts appear increases with grid-size. This is because the Nyquist $k$-values increase with the spatial sampling frequency.}
    \label{converge_test}
\end{figure}

In \fref{converge_test}, results for $M = 64$ show significant deviations compared to results from denser grids. This suggests that the grid-size is insufficiently dense to accurately represent the system. For the three largest grid sizes, \fref{converge_test}(a) and (b) show minimal variation. However, \fref{converge_test}(c) reveals a notable difference between grid sizes, as the $256^3$ and $128^3$ grid sizes exhibit ultraviolet features that appear absent in the $512^3$ grid. This feature is attributed to the Nyquist-frequency of Fourier-transforms.

Just as temporal Fourier-transforms have a maximum resolvable frequency determined by the sampling frequency, a spatial Fourier-transform has maximum resolvable $k$-values based on the spatial sampling frequency for each dimension. For a grid with $M^3$ points and lengths $L_i$, we have
\begin{equation}
\left[ \Delta x, \Delta y, \Delta z \right] = \left[ \frac{L_x}{M}, \frac{L_y}{M}, \frac{L_z}{M} \right] 
\end{equation}
        \vspace{-20pt}
\begin{equation}
k_\textrm{Nyq, i} = \frac{1}{2} \frac{2\pi}{\Delta i}, \; \; i \in \{x, y, z\}
\label{Nyq}
\end{equation}
Occupation of $k$-values beyond the Nyquist limit (Eq. \ref{Nyq}), can result in aliasing, where the condensate is misrepresented in Fourier-space. This results in the observed artifacts. Grids with more points provide higher spatial sampling, which shifts the ultraviolet artifact to higher $k$-values. 
    
For our simulations, we opted to use a grid size of $M = 256$. This allows a sufficient gap between the dissipation $k$-scale and the Nyquist $k$-scale. It also allows for efficient computation, with each second of Float64 dynamics taking approximately 38 minutes. Taking full advantage of this, we were able to test a high number of $U_F$ values [\fref{excite_energies}(e)]. 
\section{Spectral Analysis}
\subsection{Calculation of spectral densities using Plancherel's theorem}\label{Parseval_deriv}
Expressing kinetic energy in terms of our density-weighted velocity field, we can utilize Plancherel's theorem to equivalently express the integral in $k$-space rather than position-space
\begin{align}
    E_\textrm{kin}&= \frac{m}{2} \int d\textbf{r} \; |\textbf{u}(\textbf{r})|^2 = \frac{m}{2} \int d\textbf{k} \; |\Tilde{\textbf{u}}(\textbf{k})|^2,
\label{Parseval}
\end{align}
where $\Tilde{\textbf{u}}(\textbf{k})$ is the three-dimensional Fourier-transform of $\textbf{u}(\textbf{r})$. Integrating the right-hand side of Eq. \ref{Parseval} over the solid angle in $k$-space, we obtain a distribution which integrates over $k$ to give the kinetic energy.
\begin{align}
    E_\textrm{kin} & = \int^\infty_0 dk \; \frac{m k^2}{2} \int^{2\pi}_0 d\phi_k \int^{\pi}_0 d\theta_k \;\sin(\theta) |\Tilde{\textbf{u}}(k, \phi_k, \theta_k)|^2 \nonumber \\
    & = \int^\infty_0 dk \; \varepsilon_\textrm{kin}(k)
    \label{vps_def}
\end{align}
There is then a temptation to describe $\varepsilon_\textrm{kin}(k)$ as the kinetic energy spectral density. However, this is somewhat misleading, as the components of $\varepsilon_\textrm{kin}(k)$ do not sum locally in $k$-space, limiting it's interpretation as an energy density.
\begin{equation}
    \varepsilon_\textrm{kin}(k) \ne \varepsilon^i_\textrm{kin}(k) + \varepsilon^c_\textrm{kin}(k) + \varepsilon^q_\textrm{kin}(k)
\end{equation}
\vspace{3pt}
Using $\varepsilon_\textrm{kin}(k)$ also implicitly introduces a semi-classical approximation, as quantum phase information is neglected when calculating $|\Tilde{\textbf{u}}|^2$ in Eq. \ref{vps_def}. We can more appropriately describe $\varepsilon_\textrm{kin}(k)$ as a velocity power spectral density. For wavefunctions represented on a cartesian grid (typically the case in numerical studies), there will also be fewer points with small $|k|$ than large, leading to a lack of resolution in the infrared region of the resulting spectra. Conversely, the $k$-values for the reformulated spectral-density (Eq. \ref{spec_IP}) are completely decoupled from the grid representing the wavefunction, allowing for arbitrarily high resolution in $k$-space. While $k$-space resolution can be arbitrarily high, the available range is still limited by Nyquist $k$-values. 
\begin{figure}[t!]
    \centering
    \includegraphics[width=\columnwidth]{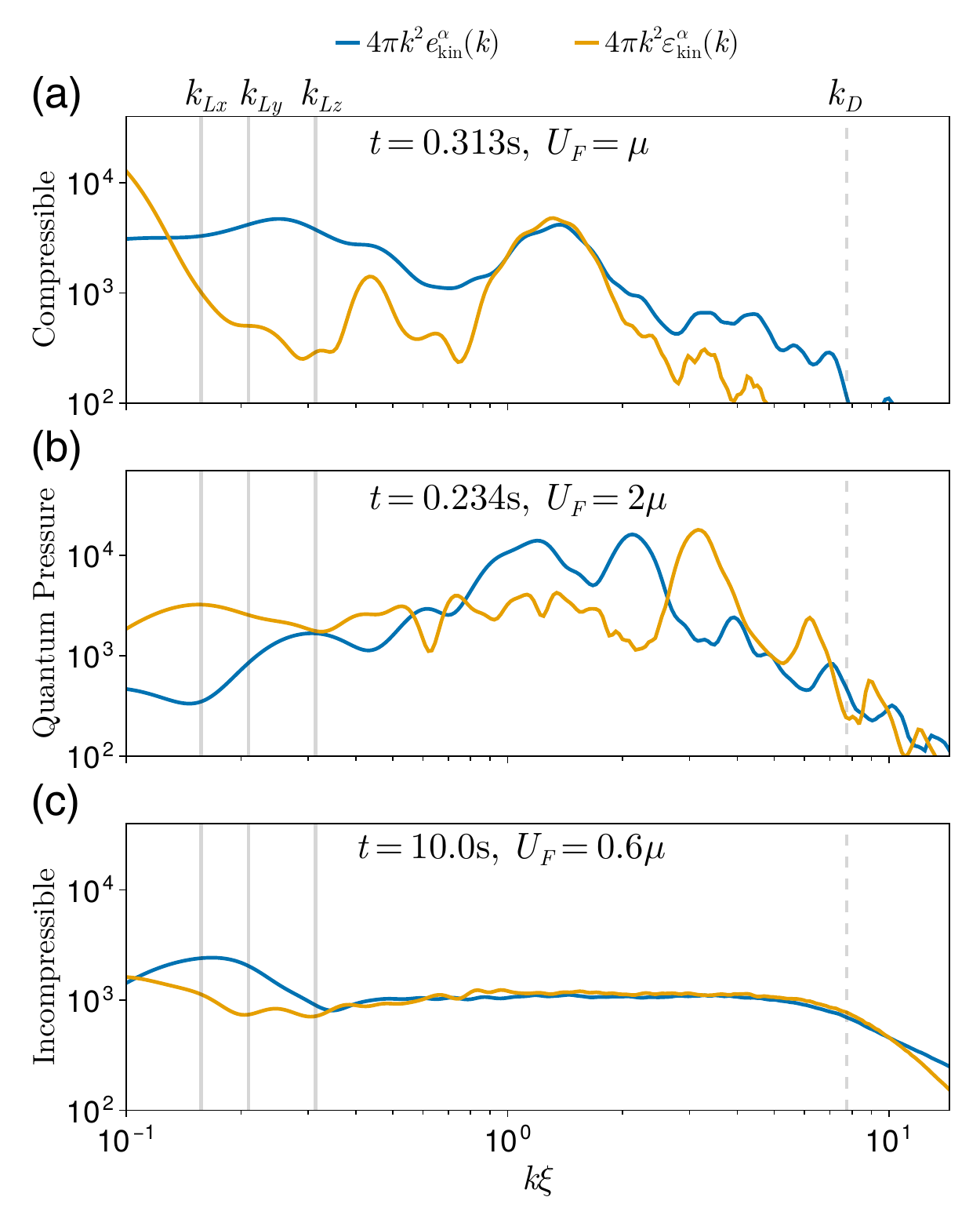}
    \caption{Comparison of velocity power spectra and energy spectral densities. (a) Compressible components. The two spectra are entirely dissimilar, except for a small central peak. \revise{(b) Quantum Pressure, where the two spectra are again distinct. (c)} Incompressible components. The two distributions show strong agreement in the inertial range, but are distinct in the ultraviolet and infrared. This suggests that the turbulence is semiclassical, corroborating the random phase approximation}
    \label{spectra_comparison}
\end{figure}
\subsection{Comparison of energy densities and velocity power spectra}\label{e_vs_e}
To demonstrate the importance of including phase information, \fref{spectra_comparison} compares equivalent energy density and velocity power spectra components. Both are calculated identically, in except that $\varepsilon^\alpha_\textrm{kin}(k)$ use the density-weighted velocity fields, and $e^\alpha_\textrm{kin}(k)$ use the $\textbf{w}^\alpha$ fields which incorporate quantum phase, defined in Eq. \ref{w_fields}. 

In \fref{spectra_comparison}, we see $\varepsilon_\textrm{kin}(k)$ and $e_\textrm{kin}(k)$ can have close agreement [\fref{spectra_comparison}(b)] or be extremely distinct [\fref{spectra_comparison}(a)] for different cases. Generally, the two distributions are similar in the inertial range for well-developed turbulence. During initial excitation and at extreme lengthscales, there is no clear pattern relating the two spectra. This indicates that for well-developed turbulence there is nontrivial phase-behaviour in the inertial range, such that neglecting quantum phase and imposing a semi-classical approximation does not introduce significant error to energy spectral densities. An interesting question then would be whether this is true for other forms of turbulence. 
        
\fref{spectra_comparison} demonstrates that inclusion of phase information is clearly important in certain cases. However, this information is typically inaccessible in experimental studies of BECs. Therefore, it would be valuable to understand the relationship between $\varepsilon_\textrm{kin}(k)$ and $e_\textrm{kin}(k)$, and identify conditions under which they are similar enough that phase information becomes less critical. 
      
\section{Comparison with Logarithmic-correction model}\label{logmodel}
Another explanation for the $k^{-3.5}$ power law is that the $k\propto k^{-3}$ stationary solution is marginally non-local~\cite{zhu_direct_2023}, requiring a log-correction of the form
\begin{align}
    n(k) &= C_d P_0^{1/3} k^{-3} \ln^{-1/3}(k/k_F).
\end{align}
The first two terms are constants and $k_F$ is the wavenumber associated with the forcing length scale, where energy is injected. 
\begin{figure}[t!]
    \centering
    \includegraphics[width=\columnwidth]{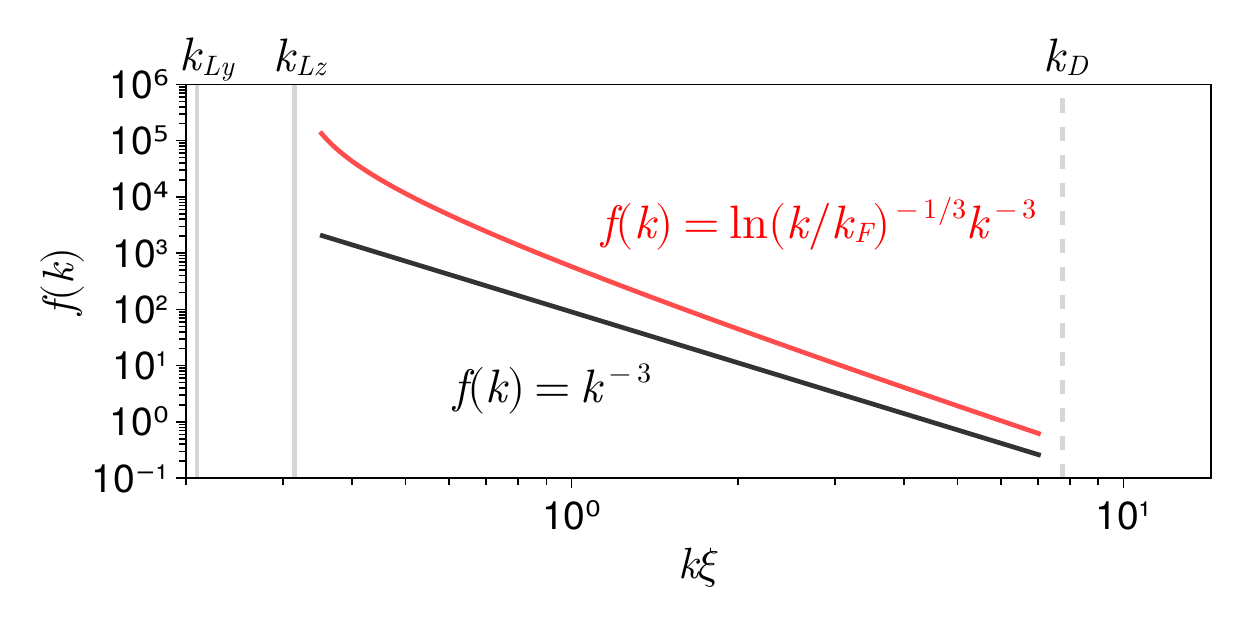}
    \caption{Log-Log plot comparing power-law (black) with a logarithmic-corrected power-law (red). If one of these theoretical models accurately described the turbulent state, we would expect the spectra to appear flat when compensated by a factor of $k^{3}$ or $\textrm{ln}(k/k_F)^{1/3}k^3$. Here $k_F = k_{L_z}$.}
    \label{logplot}
\end{figure} 

\begin{figure}[!b]
    \centering
    \includegraphics[width=\columnwidth]{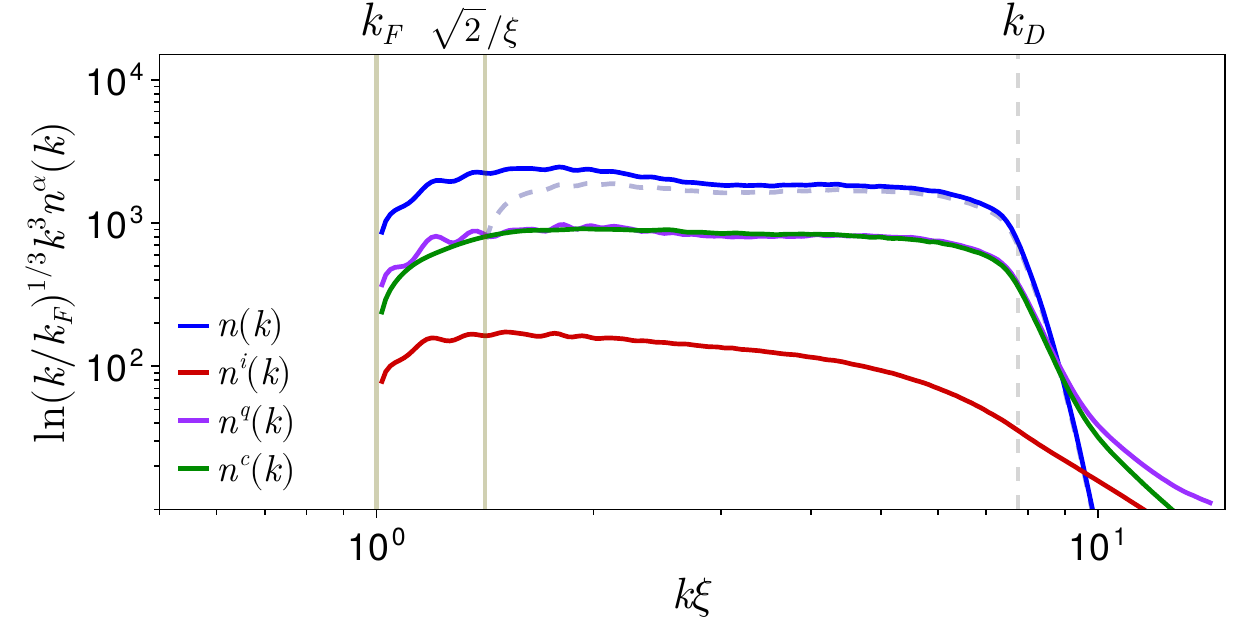}
    \caption{Momentum distribution and components compensated by logarithmic power-law for $U_F = 0.6\mu$ and the choice $k_F = 1/\xi$. The spectra appearing flat implies the logarithmic-model is accurately describing the dynamics in that particular region. While the model shows good agreement in the total $n(k)$, we see even better agreement in the compressible and quantum pressure components. The log-compensated spectrum has a narrower range of flat behavior if instead the forcing wavenumber is chosen as $\sqrt{2}/\xi$ (dashed gray line).}
    \label{zhu_comp}
\end{figure}  
This logarithmic-correction emerges from the introduction of an infrared cutoff at the forcing wavenumber, $k_F$. In Ref.~\cite{zhu_direct_2023}, $k_F$ is a defined parameter. With our method of energy injection this is not the case, and $k_F$ is slightly more ambiguous. We excite the condensate along the $z$-direction, so one might expect $k_F = k_{L_z}$. However, in practice we find the forcing peak appears at slightly higher $k$-values than $k_{L_z}$ [\fref{k3.5}], implying the forcing length scale is slightly smaller than $L_z$. Regardless, we test the model fit using a wide range of $k_F$ values.

Applying the logarithmic-correction to spectra from the $k^{-3.5}$ regime, we generally find poor agreement using $k_F$ values near the actual forcing lengthscale. However, for $k_F = 1/\xi$, \revise{corresponding to a length scale of $2\pi\xi$}, we find good agreement in the total momentum distribution and slightly stronger agreement in the compressible and quantum pressure components [\fref{zhu_comp}]. \revise{This $k_F$ value indicates a forcing scale distinct from both the obvious forcing scale $L_z$ from the drive, and the healing length. In Ref.~\cite{martirosyan_equation_2024} similar analysis showed that for 4-wave interactions the physical forcing scale due to the drive is distinct from the effective injection scale for weak wave turbulence.} 

The slightly improved fit of the model in $n^q(k)$ and $n^c(k)$, as opposed to the total momentum distribution, is likely due to incompressible excitations irrelevant to the WWT cascade being discarded. In Ref.~\cite{zhu_direct_2023} there is strong agreement with no partitioning of the spectrum, possibly because fewer vortices are nucleated. 

Where we use a finite box-trap with hard-wall boundaries, Ref.~\cite{zhu_direct_2023} uses a homogeneous condensate with periodic boundaries. Our forcing term is an emulation of experimental forcing, whereas in Ref. \cite{zhu_direct_2023}, energy is injected in $k$-space in a narrow band about at particular $k_F$ value. Our model also does not feature any mechanism for dissipation at large lengthscales. These factors may all play a role in introducing greater incompressible aspects to the steady turbulent state. 
\end{document}